\documentclass[aps,pre,floatfix,nofootinbib]{revtex4}
\usepackage{graphicx} %\usepackage{amsmath} \usepackage{amssymb}
\usepackage[sort&compress]{natbib}
\usepackage{graphicx}
\usepackage{amsmath}\usepackage{amssymb}
\usepackage{subfigure}
\usepackage{color} 

\newcommand{\BEQ}{\begin{equation}}
\newcommand{\EEQ}{\end{equation}}
\newcommand{\BEA}{\begin{eqnarray}}
\newcommand{\EEA}{\end{eqnarray}}
\newcommand{\comment}[1]{}

\begin{document}
\draft

\title{Memory-induced mechanism for self-sustaining activity in
  networks}

\author{A. E. Allahverdyan$^{1}$, G. Ver Steeg$^{2}$ and
  A. Galstyan$^{2}$}

\address{$^{1}$Yerevan Physics Institute, Alikhanian Brothers Street
2, Yerevan 375036, Armenia,\\
$^{2}$USC Information Sciences Institute, 4676 Admiralty Way, 
Marina del Rey, CA 90292, USA \\
}

\date{\today}

\begin{abstract} 

  We study a mechanism of activity sustaining on networks inspired by
  a well-known model of neuronal dynamics. Our primary focus is the
  emergence of self-sustaining collective activity patterns, where no
  single node can stay active by itself, but the activity provided
  initially is sustained within the collective of interacting
  agents. In contrast to existing models of self-sustaining activity
  that are caused by (long) loops present in the network, here we
  focus on tree--like structures and examine activation mechanisms
  that are due to temporal memory of the nodes. This approach is
  motivated by applications in social media, where long network loops
  are rare or absent. Our results suggest that under a weak behavioral
  noise, the nodes robustly split into several clusters, with partial
  synchronization of nodes within each cluster. We also study the
  randomly-weighted version of the models where the nodes are allowed
  to change their connection strength (this can model attention
  redistribution), and show that it does facilitate the self-sustained
  activity.

\end{abstract}

%\pacs{89.75.Hc, 89.65.-s, 84.35.+i, 89.20.Hh}

%% 64.60.Cn Order-disorder transformations; statistical mechanics of
%% model systems

%% 75.10.Hk Classical spin models

%% 89.75.Hc Networks and genealogical trees

%% 89.65.-s Social and economic systems

%% Self-organization complex systems, 89.75.Fb

%% Neural networks, 84.35.+i

%% Internet, 89.20.Hh

\maketitle

\section{Introduction}

The emergence of online social networking and microblogging sites has
facilitated novel channels of communication that did not exist under
the traditional centralized dissemination model \cite{lazar}. Online
platforms such as Twitter or Facebook enable users to produce content
and distribute it through social ties. This process is often modeled
via threshold elements, where a node is affected, or {\em activated},
whenever the influence of its local environment exceeds a certain
threshold.  The affected node can then propagate the influence
further, probably causing a global activation cascade throughout the
whole system \cite{watts,dodds_watts,galstyan,gleeson}. Studies of
such cascades and related dynamic processes on networks have
attracted significant attention; see \cite{porter2014,barrat} for
recent surveys.

Threshold models have a long history in quantitative sociology, and
different variants have been used for describing weak ties
\cite{weak_ties}, social impact \cite{granovetter,latane}, and
economic activity \cite{morris,durlauf,stauffer,econo,cortes}. Another
line of research where threshold models are prevalent is neuronal
systems~\cite{peto,arbib,stro,integrate_fire}. Indeed, several
scholars have noticed certain analogies between social and neuronal
systems \cite{klopf,shvyrkov,nowak,vidal,larson}. In particular, both
social agents and neurons can be viewed as information processing
units that gather information from their local neighborhood and act on
that information (firing in the case of neurons, posting in the case
of social media users). Furthermore, the electrical potential of the
neuron can be reinterpreted as an accumulated ``information'' of the
agent. Those analogies were exploited in \cite{peto}, where
conditional reflexes of a single (social) agent were described via a
neuronal model.

A popular approach for describing neuronal dynamics is the so-called
integrate-and-fire (IF) family of models \cite{stro,integrate_fire},
which emerged from the first formalization of neuronal networks
proposed by McCulloch and Pitts \cite{peto}. The IF model was used to
examine the conditions under which certain network structures exhibit
self-sustained dynamics due to small-world effects
\cite{roxin,kaski,qian}; see also \cite{bassett,davidsen} in this
context. More recently, the authors of \cite{arenas} introduced a
strongly driven, perfect-memory integrate-and-fire neuronal model for
describing the collective dynamics of social networks. The model
overcomes one of the main limitations of the cascade approaches
(absence of feedback, i.e., each node is active only once) and is
applicable to the quantitative description of Twitter data
\cite{arenas}.

In this paper we formulate a simple IF-based model to study mechanisms
of collectively sustained activity patterns in networks. Note that we
differentiate between activity cascades|activity propagating through
the whole system at least once|and the collective activity sustaining,
where an agent cannot activate in isolation, but a collective of
agents on a network does keep (possibly locally) an initial activity
over a sufficiently long time period.

There is a large body of work on collective activity patterns in
neuronal systems \cite{E-Y,abbott,schuster,anni,mcgraw,chin}.  This
activity relates to proactive functions of the brain, e.g., attention and
memory \cite{fries,pipa}.  The main mechanism for self-sustaining
collective activity in such systems is based on the existence of
cycles in networks, which effectively serve as feedback loops
\cite{arbib,mikhailov}. Such cycles, also called {\em reverberators},
are abundant even in random networks \cite{anni,mcgraw,chin}. Recent
research has focused on understanding how the loops contribute to
sustained activity and how their embedding in a network affects the
activity patterns \cite{anni,mcgraw,chin}. Besides neuronal networks,
self-sustained collective oscillations are observed in other systems
\cite{others}.

Here we are interested in alternative mechanisms for collective
activity, which do not rely on the existence of long cycles. In
particular, we are motivated by recent observations that the effective
networks in social media might not necessarily have long loops. For
instance, it has been established that the functional networks
inferred using the activity traces of the users are tree-like, with
relatively short loops; see Sec.~\ref{choice} for details. Thus, it is
important to have a different mechanism of self-sustaining cascades
that does not involve long feedback loops.

The proposed mechanism of self-sustaining activity patterns is
based on temporal memory of an individual agent. Namely, we show that
there is a parameter range of memory and inter-agent coupling that
allows self-sustaining collective activity patterns. The underlying
reason for emergence of this regime is the fragmentation of the
network into (partially synchronized) clusters. Furthermore, it is
shown that under a weak (behavioral) noise, this fragmentation is
robust with respect to details of the activation process, number of
nodes, initial conditions, the noise model, {\em etc}.

We also study the activation dynamic on randomly weighted networks,
where the weights can be interpreted as attention the nodes pay to
their neighbors.  Our numerical results indicate that the introduction
of random weights does facilitate activity sustaining, sometimes even
violating intuitively obvious bounds.

The remainder of this paper is organized as follows. We introduce the model
in Section~\ref{model}. Section \ref{strong_c} outlines the
memory-less situation, while in Section \ref{sus} we deduce the
implications of an agent's memory for self-sustaining collective
activity. In Section \ref{section_noise} we study the effect of weak
noise. Section \ref{section_weighted} shows how limitations on the
agent's attention can be described via a (randomly) weighted network. We
conclude by discussing our main results in Section
\ref{sec:conclusion}.

\section{The model}
\label{model}

\subsection{Neurons versus agents}

Before introducing our model, let us recall some essential facts about
information processing by neurons. A neuron collects signals from its
environment and from other neurons via dendrites (in-going channels)
\cite{peto}. These signals build up its electrical potential. Once
this potential is larger than a certain threshold, the neuron fires
(generates a signal) via its single out-going channel, axon
\footnote{There are primarily two types of signaling: spiking is a relatively irregular activity; bursting is intensive and regular~\cite{peto}.}. The signal is received by all neurons that are
connected (via their dendrites) to the axon. After the neuron fires,
the potential nullifies. It can accumulate again after some null
period \cite{peto}. Certain neurons do have an internal ability to
generate signals even in isolation from other neurons \cite{peto}.

To establish the analogy with a social system, we assume that the
$i$'th agent in our social system is characterized by {\em
  informational} potential $w_i$, which can be modified due to content
that the agent receives from his neighbors. We assume that whenever $w_i$
overcomes a certain threshold $u_i$, the agent displays the generated
content (fires). Immediately after that event, the information
potential nullifies, and then starts to accumulate again.

\subsection{Basic equations}

For each agent $i$ ($i=1,...,N$) we introduce a variable
$m_i(t)$ that can assume two values, $0$ (passive with respect to
content generation) and $1$ (content generating) at discrete time
$t$. The state dynamics are described as
\begin{eqnarray}
\label{n1}
m_i(t)&=&\vartheta [\, w_i(t)-u_i\,],\\
w_i(t+1)&=&[1-m_i(t)]\times \nonumber\\ 
&& [\, (1-\kappa_i)w_i(t) +r_i+{\sum}_jq_{ij}m_j(t) \, ],~~
\label{n2}
\end{eqnarray}
where $w_i(t)\geq 0$ is the information potential accumulated by the
agent $i$ till time $t$, ${\sum}_{j}q_{ij}m_j(k)$ is the cumulative
influence from other agents (we assume $q_{ii}=0$), and $\vartheta(x)$
is the step function:
\begin{eqnarray}
  \label{eq:24}
\vartheta (x< 0)=0,~~ \vartheta(x\geq 0)=1. 
\end{eqnarray}
In Eqs.~(\ref{n1}, \ref{n2}), $u_i$ is the threshold, $r_i$ is the
external rate of potential generation, while $1\geq\kappa_i\geq 0$ can
be interpreted as the rate of memory decay (or forgetting). The
prefactor $[1-m_i(t)]$ in front of Eq.~(\ref{n2}) ensures that the
information potential of an agent nullifies after displaying content.

The influence $q_{ij}m_j(t)$ of the agent $j$ on the potential $w_i$
of $i$ is non-zero provided that $j$ is in its content-displaying
state, $m_j(t)=1$. Depending on the sign of $q_{ij}$, this influence
can encourage or discourage $i$ in expressing itself. Here we assume
that demotivating agents are absent, \footnote{Facilitating and
  inhibiting types are well-known for neurons. Dale's law states
  that a neuron cannot be both inhibiting and facilitating (i.e.,
  facilitating some neurons and inhibiting different ones). The law
  has certain exclusions, but it does hold for the cortex neurons
  \cite{peto}. In the human brain, some 80 \% of neurons are facilitating
  \cite{peto}. Hence the assumption that all neurons are facilitating
  is frequently made in neuronal models \cite{arbib}. It is known
  that inhibiting neurons can lead to novel effects
  \cite{arbib,peto}.}  so we have
\begin{eqnarray}
  \label{eq:3}
  q_{ij}\geq 0, ~~~{\rm and}~~~ w_i\geq 0.
\end{eqnarray}

In the memoryless limit $\kappa=1$, Eqs.~(\ref{n1}, \ref{n2}) relate
to the deterministic neuronal dynamics model introduced by McCulloch
and Pitts \cite{peto}. The original McCulloch-Pitts model also
included strong inhibition that is absent in the present model. In the
continuous time-limit, Eqs.~(\ref{n1}, \ref{n2}) reduce to the
integrate and fire model \cite{stro,integrate_fire}.

\subsection{Isolated agent}
\label{golosa}

For $q_{ij}=0$, Eq.~(\ref{n2}) reduces to
\begin{eqnarray}
  \label{n4}
  w(0)=0, \qquad w(t)=\frac{r}{\kappa}[1-(1-\kappa)^{t}], ~~ t=1,2,...
\end{eqnarray}
Thus, $w(t)$ monotonically increases from $0$ to $\frac{r}{\kappa}$:
if $u<\frac{r}{\kappa}$, the agent fires with a period determined from
Eq.~(\ref{n4}); otherwise it never fires.

\subsection{Behavioral noise}
\label{nono}

The deterministic firing rule given by Eq.~(\ref{n1}) should be
modified to account for agents with behavioral noise. Specifically, we
want to make it possible for an agent to fire (not to fire) even
for sub-threshold (super-threshold) values of the potential.  The
noise will be implemented by assuming that the threshold $u_i+v_i(t)$
has (besides the deterministic component $u_i$) a random component
$v_i(t)$. These quantities are independently distributed over $t$ and
$i$. We shall employ two models for the behavior of $v_i(t)$.

Within the first model, $v_i(t)$ is a trichotomic random variable,
which takes values $v_i(t)=\pm V$ with probabilities $\frac{\eta}{2}$
each, and $v_i(t)=0$ (no noise) with probability $1-\eta$. In this
representation, $\eta$ describes the magnitude of the noise. We assume
that $V$ is a large number, so that with probability $\eta$, the agent
ignores $w_i$ and activates (or does not activate) randomly. Thus,
instead of Eq.~(\ref{n1}), we now have
\begin{eqnarray}
  \label{noise}
m_i(t)=\vartheta [\,\phi_i(t)(\, w_i(t)-u_i\,)\,].
\end{eqnarray}
In Eq.~(\ref{n1}), $\phi_i$ are independent (over $i$ and over $t$) random
variables that assume values $\pm 1$ with probabilities
\begin{eqnarray}
  \label{n33}
  {\rm Pr}[\phi_i=1]=1-\eta, \qquad   {\rm Pr}[\phi_i=-1]=\eta.
\end{eqnarray}

Our second model, which has wide applicability in neural network
literature \cite{peto}, amounts to replacing the step function by a
sigmoid function. Instead of Eq.~(\ref{n1}), we now have
\begin{eqnarray}
  \label{noise_temperature}
  m_i(t)&=& 1 ~~~{\rm with\, probability}~~~ \frac{e^{(
      w_i(t)-u_i\,)/\theta}}{1+e^{(w_i(t)-u_i\,)/\theta}}, \\ 
&=& 0 ~~~{\rm with\, probability}~~~
\frac{1}{1+e^{(w_i(t)-u_i\,)/\theta}}, 
\end{eqnarray}
where $\theta\geq 0$ has the (formal) meaning of temperature, so that
the noiseless model is recovered when $\theta\to 0$.

In this paper we limit ourselves to the weak-noise limit ($\eta$ and
$\theta$ are small). As illustrated below, both models (\ref{noise},
\ref{noise_temperature}) lead to similar predictions.

\subsection{Choice of the network}
\label{choice}

To proceed further, we have to specify the network structure on which
the activation process unfolds.  For social media platforms such as Twitter, it
seems straightforward to identify the network structure using the
declared list of friends and followers. This choice, however, does not
necessarily reflect the true interactions that are responsible for the
activity spreading~\cite{huberman}. Indeed, there is accumulating evidence
that a considerable part of links in Twitter's follower/friendship
network are meaningless, in the sense that they do not necessarily
participate in information diffusion. Instead, one has to pay
attention to {\em functional} links, which are based on the observed
activity patterns of the users. One possible way for inferring such
functional networks based on transfer entropy (or information
transfer) was suggested in Ref.~\cite{www2012}.  As compared to
formally declared networks, the ones based on the activity dynamics
have fewer loops~\cite{huberman,www2012}. 

Here is an example that illustrates this point and motivates the
introduction of a model network below. A dataset was collected from
Spanish Twitter, which contains 0.5 million Spanish messages publicly
exchanged through this platform from the 25th of April to the 25th of
May, 2011. This dataset was already studied in \cite{arenas}. We
constructed the user-activity network via the transfer-entropy method
\cite{www2012,wsdm2013}, and found that over 99 \% of all nodes
(users) do have loops attached to them; among those loops 84 \% have
length 2 (reciprocity), while other loops have length 3; loops of
length $\geq$4 are either negligible or absent.

\begin{figure*}[htbp]
\section*{\large Figures}
    \includegraphics[width=0.6\columnwidth]{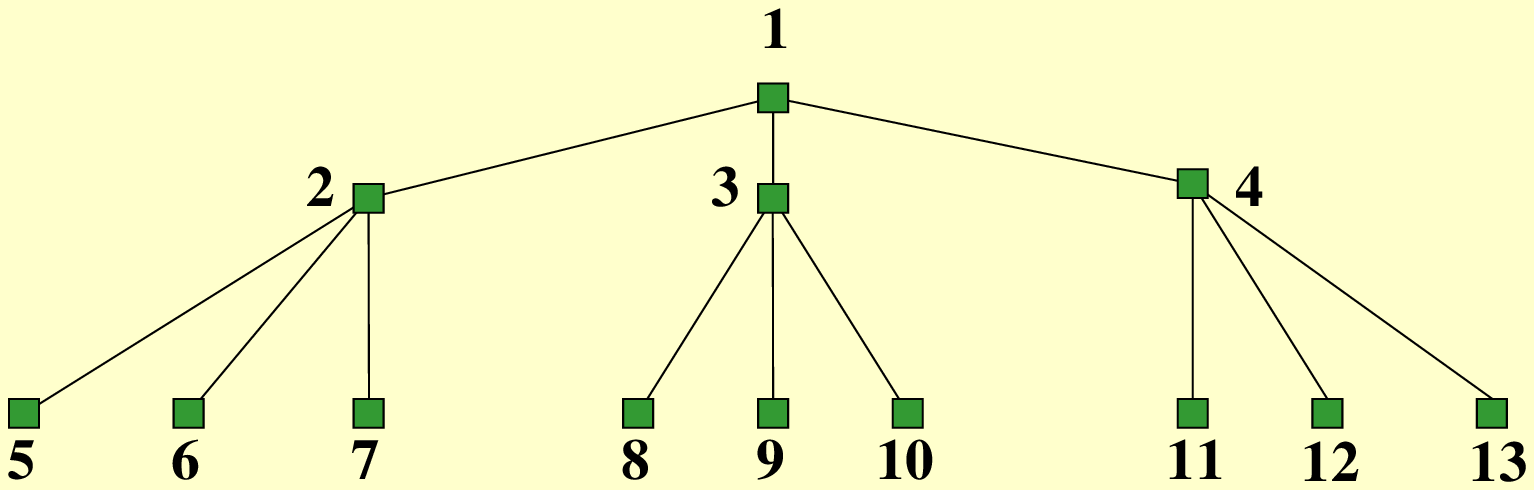} 
\caption{(Color online)
The Cayley tree with $K=3$ and two generations. The nodes
  $5-13$ belong to periphery. The nodes are
      numbered from the root of the tree, i.e. $k=1$ is the root,
      $k=g$ is the last node of the core, and $k=N$ is the last
      node in the periphery; cf. Eq.~(\ref{eq:7}).}
\label{fig00}
\end{figure*}

Hence, we need a network model that is (nearly) symmetric, does not
contain loops of length 4 and larger and allows a well-controlled
introduction of triangles (loops of length 3). As the basic model, here we
focus on the (undirected) Cayley tree: one node is taken as the root
(zeroth generation) and then $K$ symmetric links are drawn from it
(nodes of the first generation); see Fig.~\ref{fig00}. In subsequent
generations, each node of the previous generation produces $K$
symmetric links.  Nodes of the last generation belong to the
periphery, and each of them has only one connection. All other nodes belong to
the core. The root is connected with $K$ nodes, while all other nodes
in the core have $K+1$ connections. If the Cayley tree has $p$
generations, then the overall number of nodes is
\begin{eqnarray}
  \label{eq:7}
N= (K^{p+1}-1)/(K-1) =gK+1,
\end{eqnarray}
where $g$ is the number of nodes in the core. 

Obviously, a Cayley tree has no loops. In our analysis, we will also
examine the effect of short loops via addition of a fixed amount of
triangles, generated by randomly linking pairs of nodes from the same
generation.  A similar procedure of studying the influence of
triangles (or other mesoscopic network structures) is frequently
applied in literature \cite{peron}.  As we show below, these triangles
do not lead to quantitative changes, at least for the scenarios
examined here; see \cite{peron} for related results.

In addition to being loop-free, the Cayley tree has another
characteristic that makes it a suitable choice for the present
study. Specifically, one of the main mesoscopic motifs observed in
real-world complex networks is the {\em core-periphery structure},
where some of the nodes are well interconnected (core), while
sparsely connected periphery is around the core; see
\cite{freeman,everett,bocca,puck} for reviews. The Cayley tree is one of
the simplest network models that exhibits features consistent with the
definition of the core-periphery structure \cite{freeman}: {\it (i)}
the core is more connected than the periphery; {\it (ii)} core nodes
are located on geodesics connecting many pairs of nodes (large
betweenness degree); {\it (iii)} core nodes are minimally distant from
(possibly many) other nodes. As we show below, the identification of
the core-periphery structure for the Cayley tree is reflected in the
dynamics of the model defined on it.

\section{Clustering and synchronization due to strong coupling}
\label{strong_c}

In Eqs.~(\ref{n1}, \ref{n2}) we assume that all agents are equivalent
and that the network is not weighted:
\begin{eqnarray}
  \label{equip}
 r=r_i, ~~~~ u=u_i, ~~~~ \kappa=\kappa_i , ~~~~ q_{ij}=q.
  \label{equiv} 
\end{eqnarray}
Note that for
\begin{eqnarray}
  \label{eq:6}
  q>u-r>0,
\end{eqnarray}
activation of one agent is sufficient for exciting his
neighbors. This is not a realistic condition. Below we will almost always
assume that $q<u-r$. However, let us first discuss what happens
when Eq.~(\ref{eq:6}) is valid. We can focus on the memoryless
situation $\kappa=1$, since under the condition Eq.(\ref{eq:6}), the memory
is not relevant. We stress that $u>r$ in Eq.~(\ref{eq:6}) means that
that no agent is able to activate spontaneously by itself (i.e., for $q=0$);
cf. Eq.~(\ref{n4}).

Our numerical results are summarized as follows. Provided that there
is at least one active agent ($m_i(0)=1$) initially, after a few time-steps the system converges to a fully synchronized state, where it
separates into two clusters: the first cluster mainly includes the
core, while the second cluster mainly includes periphery. If the first
cluster is active, then the second one is passive, and {\it vice
  versa}. All agents have nearly the same firing frequency that is
close to its maximal value of $0.5$. Finally, this scenario exists
for any $K\geq 1$.

Thus, the clusters alternate their active states, with one cluster
collectively {\em exciting} the other. Note that neither of those
clusters can sustain activity by itself, because after each
activation, the potential for all the nodes drops to zero, and no
isolated agent can be activated by itself.

If instead of Eq.~(\ref{eq:6}) we set $u>r+q$ (but $r<u$ and
$\kappa=1$), then the system converges to a trivial passive fixed
point for all initial states, where any initial activity dies out
after a few time-steps. This behavior crucially depends on the fact
that the Cayley tree does not have sufficiently long loops
\footnote{If the network contains a sufficient number of mesoscopic loops,
  the self-sustaining activity in the memoryless situation $\kappa=1$
  can be non-trivial; see \cite{mcgraw,chin} for examples. For
  instance, consider Eqs.~(\ref{n1}, \ref{n2}) with $\kappa=1$ on the
  directed Erdos-Renyi network: for each agent $i$ one randomly
  selects $K$ other agents and assigns to them $K$ connections with
  equal weights $q_{ij}=q$. This network has a rich structure of
  mesoscopic (length $\sim \ln N$) loops. In particular, the above
  two-cluster synchronization can also exist when condition
  (\ref{eq:6}) does not hold (i.e., when $u>r+q$, $r<u$ and
  $\kappa=1$), but provided that $K$ (the average number of neighbors)
  is sufficiently large. Here are examples of the synchronization
  threshold in that situation:
  \begin{eqnarray}
    \label{eq:13}
&& q>u-r ~~ {\rm for} ~~ 3\geq K\geq 2, \nonumber \\
&& q>(u-r)/2 ~~ {\rm for} ~~ 6\geq K\geq 4, \nonumber \\
&& q>(u-r)/3 ~~ {\rm for} ~~ 9\geq K\geq 7, \nonumber \\
&& q>(u-r)/4 ~~ {\rm for} ~~ 11\geq K\geq 10, ~~{\it etc}.\nonumber
  \end{eqnarray}
}.

The above results remain qualitatively valid after adding (randomly) up to
$\sim N/K\approx g$ triangles. The same threshold (\ref{eq:6}) holds
in this case. 

\section{Collective dynamics driven by memory}
\label{sus}

\subsection{Thresholds of self-sustaining activity}

If the coupling is not strong, $0<q<u-r$, our numerical results suggest 
that there is a self-sustaining collective activity regime, which exists
under the following conditions:\\
$\imath)$ There should be sufficiently many neighbors: 
\begin{eqnarray}
  \label{eq:9}
K\geq 2.
\end{eqnarray} \\
$\imath \imath)$ The memory should be non-zero, $\kappa<1$, but still small enough
so that no agent can fire in isolation [see Eq.~(\ref{n4}), and note
that conditions of Eq.(\ref{equip}) hold]:
\begin{eqnarray}
  \label{eq:8}
u> r/\kappa.
\end{eqnarray}
Without loss of generality, below we set
\begin{eqnarray}
  \label{eq:4}
  u_i=1,
\end{eqnarray}
Note that $u=u_i$ can be adjusted by varying the parameters $1-\kappa$,
$r$ and $q$; see Eqs.~(\ref{n1}, \ref{n2}).

The simulations of the model reveal that the collective activity
pattern in this regime|and its very existence|depend on initial
conditions $\{w_i(0)\}_{i=1}^N$ of Eqs.~(\ref{n1}, \ref{n2}), i.e., the
choice of initially active nodes. To describe this dependence, we
assume that in the initial vector $\{w_i(0)\}_{i=1}^N$ each $w_i(0)$
is generated randomly and homogeneously in the interval.

\begin{eqnarray}
  \label{eq:29}
  [0,b], ~~~ 2>b>1.
\end{eqnarray}
For $N\gg 1$, Eqs.~(\ref{eq:4}, \ref{eq:29}) imply that $\simeq
(b-1)N/b$ agents are active in the initial state (for them
$m_i(0)=1$).
 
Numerical results show that there exists an upper threshold ${\cal Q}^+
<1$, such that for $q>{\cal Q}^+$ (and all other parameters being fixed),
the collective activity is sustained in time for (almost) all
realizations of the random initial state:
\begin{eqnarray}
  \label{th1}
q>{\cal Q}^+ \Longrightarrow m(t)\equiv 
\frac{1}{N}\sum_{k=1}^N m_i(t)
\not =0 ~~ {\rm for}~~ t>0.
\end{eqnarray}
Likewise, there is a lower threshold ${\cal Q}^-<{\cal Q}^+$ such that
for $q<{\cal Q}^-$ the activity decays to zero in a finite time $T$
for (almost) all initial states:
\begin{eqnarray}
  \label{th2}
q<{\cal Q}^- \Longrightarrow m(t)=0 ~~ {\rm for}~~ t>T.
\end{eqnarray}
In the range ${\cal Q}^- \leq q\leq {\cal Q}^+$, whether or not the
initial activity will be sustained depends on the realization of the
random initial state $\{w_i(0)\}_{i=1}^N$; see Figs.~\ref{f1},
\ref{f11}, \ref{f2} and \ref{f3}.

The thresholds ${\cal Q}^+$ and ${\cal Q}^-$ depend on
all the model parameters: $\kappa$, $r$, $b$, $K$ and $N$. They are
decreasing functions of $N$, provided $N$ is sufficiently large; see
Table I. They are also increasing functions of $\kappa$ and saturate at
${\cal Q}^+={\cal Q}^-=u=1$ for $\kappa=1$ (no memory); see Section
\ref{strong_c}. The difference ${\cal Q}^+ - {\cal Q}^-$ is
small, but it persists for large values of $N$; see Table I.

Consider a pertinent case where only one agent is active initially. We
find that in this case the thresholds are still non-trivial; e.g., for
parameters of Fig.~\ref{f3_a} we obtain ${\cal Q}^+[b\to 1+] =0.524$,
while ${\cal Q}^+[b=1.8]=0.358$. Also, for this example of $b\to 1$,
the collective activity $m(t)=\frac{1}{N}\sum_{k=1}^N m_i(t)$ reaches
$0.1$, when starting from $1/N=1.66\times 10^{-4}$ at $t=0$.

Note that there is a simple lower bound on ${\cal Q}^-$:
\begin{eqnarray}
  \label{bo}
{\cal Q}^-> \frac{1-r}{K}.
\end{eqnarray}
To understand the origin of this bound, recall that owing to
Eq.~(\ref{eq:8}), the agents cannot activate spontaneously. If
$q<\frac{1-r}{K}$, then even all stimulating connections (acting
together) cannot activate the agent, so that no initial activity can
be sustained.

Finally, note that for cascading processes there is typically a single
threshold that depends on network structure \cite{porter2014,barrat},
even if one takes into account possible feedback effects in the
activation dynamics \cite{arenas}.

\subsection{Clustering and synchronization}

\begin{figure*}[htbp]
     \label{f1_a}
    \includegraphics[width=0.6\columnwidth]{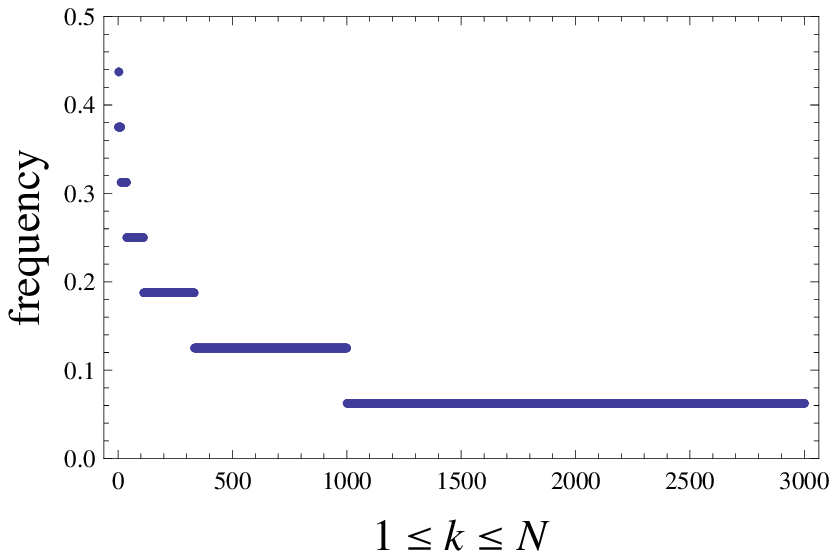} 
    \caption{(Color online) Firing frequency and activity for agents
      with memory in the noiseless situation. The agents are numbered
      as in Fig.~\ref{fig00}. \\
      Parameters: $u=1$, $r=0.1$, $\kappa=0.101$ and $q=0.36$. This
      value of $q$ is between of two thresholds ${\cal Q}^+>q>{\cal
        Q}^-$; see Table I. The initial condition was generated with
      $b=1.8$; see Eq.~(\ref{eq:29}). Parameters of the Cayley tree:
      $K=3$, $N=3\times 10^3+1$ (total number of nodes). \\
      Frequency of each node versus the number of node number.
      The frequency is defined as the number of firings in the
      time-interval $[1200,1600]$ divided over the interval length
      $400$. For this specific realization of the initial state, the
      cluster structure is very regular.   }
\label{f1}
\end{figure*}

\begin{figure*}[htbp]
\centering
    \subfigure[]{
     \label{f1_b}
    \includegraphics[width=0.6\columnwidth]{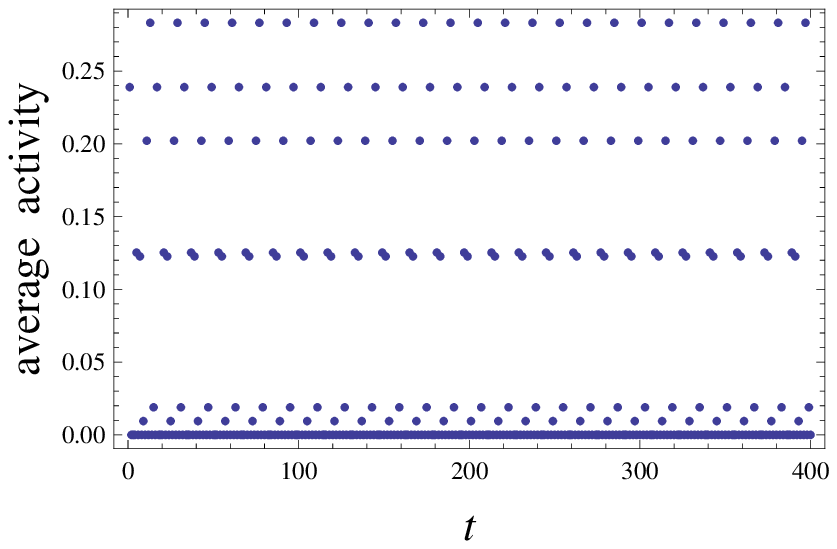} 
    }
    \subfigure[]{
     \label{f1_c}
   \includegraphics[width=0.6\columnwidth]{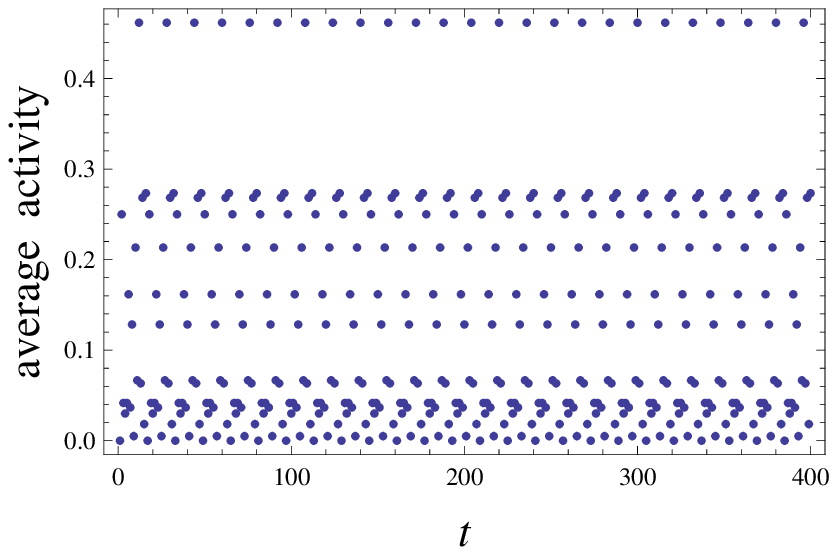} 
    }
    \caption{(Color online) The same parameters as in Fig.~\ref{f1_a}.\\
      (a) The average activity
      $m(t)=\frac{1}{2000}\sum_{k=1000}^{3000} m_k(t)$
      of the peripheric cluster versus time $t+1200$.\\
      (b) The average activity $m(t)=\frac{1}{700}\sum_{k=300}^{1000}
      m_k(t)$ of the outer core cluster versus time $t+1200$.  }
\label{f11}
\end{figure*}

Our analysis suggests that the activity sustaining occurs via
separation of the network into clusters (groups) of neighboring
agents, having the same firing frequency; see Figs.~\ref{f1},
\ref{f11} and \ref{f2} (note that certain nodes do not belong to any
cluster). The number of clusters and their distribution depend on the
realization of the (random) initial state; cf. Figs.~\ref{f1},
\ref{f11} with Figs.~\ref{f2}. Furthermore, this dependence persists
in the long time limit, suggesting that the long-term behavior cannot
be adequately described by mean-field analysis that involves only few
parameters. Concrete scenarios of clusterization can differ from each
other widely; e.g., there is a scenario (related to specific initial
states) where the distribution of clusters is very regular, ranging
from the core to the periphery of the Cayley tree; see
Figs.~\ref{f1_a}.

Within each cluster $C$ the firing frequency of agents is
synchronized; i.e., the collective intra-cluster activity
\begin{eqnarray}
  \label{intra}
  m_{C}(t)=\frac{1}{N_{C}}\sum_{i\in {C}}m_i(t),
\end{eqnarray}
where $N_{\rm C}$ is the number of agents in the cluster ${C}$
has a regular time-dependence; see Figs.~\ref{f1_b}, \ref{f1_c},
and \ref{f2}. This dependence is displayed via few
horizontal lines along which $m_{ C}(t)$ changes. 

The distribution of clusters shows two common features: there is a
cluster that involves the major part of the periphery and a smaller
cluster that is located inside of the core, close to the
core-periphery boundary; see Figs.~\ref{f1_a}, \ref{f2}. For all
realizations of the initial state, where the activity is sustained
over a long time, there is a clear transition between the core and the
periphery.

There are two well-separated time-scales here: after the first
time-scale ($\sim 50-100$ time steps for parameters of Figs.~\ref{f1},
\ref{f11} and \ref{f2}) the structure of clusters is visible, but the
intra-cluster activity is not synchronized; i.e., $ m_{\cal C}(t)$
looks like a random function of time. The achievement of intra-cluster
synchronization takes a much longer time, e.g., $\sim 10^3$ time-steps
for parameters of Figs.~\ref{f1}, \ref{f11} and \ref{f2}.

Figs.~\ref{f3_a} and \ref{f3_b} show that after adding (randomly
distributed) triangles (loops of length 3), the distribution of firing
frequencies becomes smeared, and the separation into well-defined
clusters is less visible. There is also a visible maximum of the
frequency distribution at the boundary between the core and periphery
(i.e., at $k\sim g$). Note that adding triangles does not change the
thresholds ${\cal Q}^+$ and ${\cal Q}^-$ if the number of triangles
is not large (smaller than 30 - 40 \% of nodes); cf. Section
\ref{choice}.

\begin{table}[ht]
  \caption{Activity sustaining 
    thresholds ${\cal Q}^+$, ${\cal Q}^-$ versus the overall
    number $N$ of agents for $u=1$, $r=0.1$, 
    $\kappa=0.101$, $K=3$ and $b=1.8$. Due to numerical errors, the
    numeric results given below overestimate (underestimate) the 
    value of ${\cal Q}^+$ (${\cal Q}^-$).
  }
\begin{tabular}{|c|c|c|}
\hline
$N$ & ${\cal Q}^+$ & ${\cal Q}^-$ \\
\hline\hline
$1.5 \times 10^3$ & 0.404 &  0.356\\
\hline
$3 \times 10^3$ & 0.365 &  0.354\\
\hline
$6 \times 10^3$ & 0.358 &  0.333\\
\hline
$1.5 \times 10^4$ & $0.358$ &  0.305\\
\hline
\end{tabular}
\end{table}

To summarize this section, the model studied here has revealed two
novel aspects of collective dynamics on networks. First, it shows that
agent memory can facilitate collective activity sustaining (via
clustering and synchronization), even on networks without long
loops. And second, the model has two relevant (upper and lower)
thresholds ${\cal Q}^+$ and ${\cal Q}^-$, so that in-between these
thresholds, the dynamics are very complex and sensitive to initial
conditions. Further generalizations of this model that include effects
of external noise and network weighting are studied below.

\subsection{Relation to prior work}
Synchronization is well-known in neuroscience, because it relates to
the normal activity pattern of brain neurons; see
\cite{peto,mikhailov,mimi} for reviews. It is seen on EEG measurement,
where macroscopic (many-neuron) activity is recorded and temporal
correlations are found between, e.g., the left and right hemispheres of
the brain \cite{peto,mikhailov,mimi}. Synchronized states are relevant
for the brain functioning; they relate to consciousness, attention,
memory, and also to pathologies (e.g., epilepsy).

For the fully connected|all neurons couple with each other with the
same weight|integrate and fire model (\ref{n1}, \ref{n2}) the fully
synchronized state was studied in \cite{stro}.  The existence of
a few-cluster synchronized state was predicted in
\cite{abbott,schuster}, though the existence of two different
thresholds ${\cal Q}^+$ and ${\cal Q}^-$ were not seen, and the
relation between memory, loop-structure of the network and the
activity sustaining was not recognized.

The existence of non-ergodic (initial state-dependent) fragmentation
into clusters is well-known for coupled chaotic maps
\cite{kaneko,crisanti,manrubia_non_ergo,mikhailov,mimi,gade}. This
many-attractor situation can be considered as a model for neuronal
memory \cite{peto}. However, the existing scenarios of clusterization
are based on a sufficiently strong inter-neuron coupling on a network
that admits long loops; cf. also
\cite{roxin,kaski,qian}. Ref.~\cite{gade} studied a specific model of
coupled maps on the Cayley tree, but only two relatively simple
synchronization scenarios were uncovered.

Thus, two novelties of our results are that we single out a specific
mechanism for generating clustering and synchronization (memory on a
network with long loops), and we show that these phenomena (apart of
various details) relate to two different thresholds ${\cal Q}^+$ and
${\cal Q}^-$. Further generalizations of this model that include
effects of external noise and network weighting are studied below.

\section{Noise-driven clusterization}
\label{section_noise}

We now consider the model Eqs. (\ref{n1}, \ref{n2}) under the
behavioral noise, defined by Eqs.~(\ref{noise}, \ref{n33}). As in the
noiseless case, we assume that both Eq.~(\ref{eq:9}) and
Eq.~(\ref{eq:8}) hold, e.g., each node has sufficiently many
neighbors, each node has sufficiently strong memory, but without the
possibility of activating spontaneously.

In addition, we will also assume that the following conditions hold:

$\imath)$ Weak noise: $\eta\ll 1$. Thus an isolated agent (for $q=0$)
will fire randomly with the average activity $\eta$.

$\imath \imath)$ Sub-threshold coupling: $0<q<{\cal Q}^-$, i.e., no
activity is sustained without the noise; see Eqs.~(\ref{th1},
\ref{th2}).

Under the above conditions, the dynamics lead to the fragmentation of
the network into several clusters; see Fig.~\ref{fig4}. A cluster is a
set of neighbor agents with approximately identical firing
frequencies; see Fig.~\ref{fig4}. The firing frequency is defined as
the number of firings in an interval
\begin{eqnarray}
  \label{freq}
[\tau_0, \tau_0 +\tau]  
\end{eqnarray}
divided over the interval length $\tau$. Here $\tau_0$ has to be
sufficiently large for the system to become independent from the
initial state $\{w_i(0)\}_{i=1}^N$. If $\tau$ is sufficiently large as
well, the distribution of frequencies does not change upon its further
increase; see Fig.~\ref{fig4}.

The overall number of clusters is never larger than 4| see
Fig.~\ref{fig4}, where it equals 4. The number of clusters decreases
for larger $q$, because clusters located in the core tend to merge
with each other: for $q$ close to (but smaller than) $u=1$, there
remain only 2 clusters|those involving (respectively) core and
periphery. The distribution of clusters does not depend on the
  initial conditions, i.e. no special conditions such as one specified
  by Eq.~(\ref{eq:29}) are needed. This is in stark contrast with the
  noiseless situation, where the number and structure of clusters are
  sensitive to initial conditions.

The smallest cluster includes the root of the Cayley tree, and has the
largest firing frequency. The largest cluster includes the whole
periphery; here the (average) firing frequency is the smallest one,
but it is still clearly larger than the noise magnitude $\eta$.

Additional structure of clusters is observed by analyzing the
short-time activity, i.e., when $\tau$ in Eq.~(\ref{freq}) is not
very large; see Fig.~\ref{fig4_1}. Each cluster consists of
sub-clusters that also have (nearly) the same firing
frequency. Comparing Figs.~\ref{fig4} and \ref{fig4_1} we see that
there are ``boundary'' agents: within the short-time [long-time]
activity they do not [do] belong to a definite cluster; cf.
Fig.~\ref{fig4_1} with Fig.~\ref{fig4}.

The above clusterization  disappears under sufficiently strong noise, which 
makes all the nodes equivalent. Fig.~\ref{fig4} shows that the
influence of weak noise is non-additive: the resulting agent activity
(even in the periphery) is larger than the noise magnitude $\eta$. The
fragmentation into clusters and the non-additivity disappear if the
magnitude of memory decreases; e.g., for parameters of Fig.~\ref{fig4}
(where $\kappa=0.101$) both effects disappear for $\kappa=0.2$.

Note that the correspondence between noisy and noiseless results is
there for limited times and for $\eta$ being sufficiently
smaller than $0.01$; see Fig.~\ref{fig4_3} for an example.

The results above were obtained under the noise model (\ref{noise},
\ref{n33}). Similar results are obtained for the model
(\ref{noise_temperature}); cf. Fig.~\ref{fig4_1} with
\ref{fig4_2}. The main difference between the two models
(provided that we identify them via $\eta=\theta$) is that for
Eq.~(\ref{noise_temperature}) the spread around the average frequency
is smaller. This is expected, since Eqs.~(\ref{noise}, \ref{n33})
refer to a noise that can assume large values. 

Finally, we note that introducing triangles does not alter the cluster
structure, but can increase the activity|sometimes sizeably; see
Fig.~\ref{fig7}.

\section{Randomly weighted network}
\label{section_weighted}

\subsection{Weighting and attention distribution}

So far, we assumed that each agent stimulates its neighbors in exactly the
same way, so that the link wights $q_{ij}$ do not depend on $j$, and
each connection gives the same contribution to the information
potential $w_i$ of the  $i$'th agent; cf. Eq.~(\ref{equip}). This is
clearly an oversimplification, as different connections are usually
prioritized differently. This prioritization can be related to the {\em
  attention} an agent pays to his neighbors\cite{v_at,hu_at}.

Below we extend our model to account for weighted attention
mechanims. We consider two different weighting schemes, {\em frozen}
(or {\em quenched}), where the link weights are static random
variables, and {\em annealed}, where the weights ate frequently
resampled from some distribution \footnote{The distinction between
  quenched (slow) and annealed (fast) disorder is well-known in
  statistical physics \cite{binder}. Recently, it was also studied in
  the context of neuronal dynamics as modeled by continuous-time
  integrate-and-fire neurons~\cite{touboul,motter}.}.

\subsection{Frozen versus annealed random weights}

Under this model, the link weights $q_{ij}$-s are frozen
(i.e., time-independent) random variables, sampled from some
distribution. To account for limited attention of the agents, we
require the cumulative weight stimulating each agent to be
fixed. Thus, we use the following weighting scheme:
\begin{eqnarray}
  \label{eq:38}
q_{ij}=q \phi_i \tau_{ij}, \qquad \sum_{j=1}^{\phi_i} \tau_{ij}=1,
\end{eqnarray}
where $\phi_i$ is the number of neighbors for agent $i$: $\phi_1=K$,
$\phi_{1<i\leq g}=K+1$, $\phi_{i>g}=1$; cf. Fig.~\ref{fig00}. 

Eq.~(\ref{eq:38}) allows a comparison with the non-weighted situation,
where $\tau_{ij}=1/\phi_i$ for all $i$ and $j$. Generally speaking, we
can generate $\tau_{ij}$-s by sampling from any distribution defined
over a $\phi_i$-dimensional simplex. Here we construct $\tau_{ij}$ by
sampling $\phi_i$ independent random numbers $n_{ij}$ and then
normalizing them \cite{malsburg}
\begin{eqnarray}
  \label{do}
\tau_{ij}=n_{ij}\left/\sum_{l=1}^{\phi_i} n_{il}\right. .
\end{eqnarray}
In the numerical results reported below, we assume that $n_{ij}$ are
generated independently and homogeneously in the interval $[0,b']$
with $b'=10$. We found that if $b'$ is sufficiently larger than $1$,
its concrete value is not essential, e.g., $b'=10$ and $b'=100$ produce
nearly identical results.

Our numerical results suggest that the introduction of (frozen) random
weights results in modified upper and lower thresholds ${\cal Q}^+_{f}$
and ${\cal Q}^-_{f}$; cf. Eqs.~(\ref{th1}, \ref{th2}). For $q>{\cal
  Q}^+_{f}$ [$q<{\cal Q}^-_{f}$], any initial activity persists [does
not persist] in the long-time limit, irrespective of the initial
conditions $\{w_i(0)\}_{i=1}^N$ and the attention distribution
$\{\tau_{ij}\}$. For ${\cal Q}^-_{f}<q<{\cal Q}^+_{f}$, the existence
of a long-time activity depends on realizations of $\{\tau_{ij}\}$ and
of $\{w_i(0)\}_{i=1}^N$.

Our analysis yields [cf. Eqs.~(\ref{th1}, \ref{th2})]
\begin{eqnarray}
  \label{eq:1}
  {\cal Q}^+_{f} < {\cal Q}^+, ~~~ 
  {\cal Q}^-_{f} < {\cal Q}^-,
\end{eqnarray}
which means that introduction of frozen weights facilitates the
long-term activity sustaining. Moreover, we find that ${\cal Q}^-_{f}$
can be lower than the trivial bound given by Eq.~(\ref{bo}), e.g., for
parameters of Fig.~\ref{fig_new_44}: ${\cal Q}^-_{f}=0.265$, whereas
(\ref{bo}) gives $0.3$. Recall that Eq.~(\ref{bo}) holds strictly only
for the non-weighted scenario. Though $q$ in Eq.~(\ref{bo}) still
characterizes the average magnitude of each connection also in the
weighted situation [cf. Eq.~(\ref{eq:38})], numerical results reported
in Figs.~\ref{fig_new_1}, \ref{fig_new_2} and \ref{fig_new_44} show
that Eq.~(\ref{bo}) does not extend to this situation (both for frozen
and annealed weighting schemes, as seen below) \footnote{ The
  introduction of frozen weights facilitates activity sustaining in
  the memoryless ($\kappa=1$) situation as well. Here the two-cluster
  synchronized state described in Section \ref{strong_c} exists even
  for $q<u-r$, i.e., below the threshold (\ref{eq:6}). The periphery
  is completely passive (peripheric agents have only one connection;
  hence they are not affected by the introduction of weights).}.

Random frozen weights increase activity (as compared to the
non-weighted situation), because there are $\tau_{ij}$ which is close
to one; hence there are links with the weight close to $Kq$, which is
$K$ times larger than for the non-weighted case, where all weights are
equal to $q$. Of course, there are also $\tau_{ij}$'s which are close
to zero, since both situations have the same average weight per node;
see Eq.~(\ref{eq:38}). But the influence of those weak weights on the
activity sustaining appears to be weaker.

Similar to the non-weighted (and noisy) scenario in Section
\ref{section_noise}, the dynamical system defined on the weighted
network factorizes into several clusters, which themselves consist of
sub-clusters, as seen by looking at the short-time activity; see
Figs.~\ref{fig_new_1} and \ref{fig_new_2}.  However, the cluster
structure is somewhat different compared to the non-weighted
situation. Namely, different clusters can have agents that fire with
(approximately) equal frequencies, but different clusters are
separated from each other by regions of low (but non-zero)
activity. This is especially visible near the threshold ${\cal
  Q}^-_{f}$; see Fig.~\ref{fig_new_44}.

For the annealed random situation
a new set of random weights is generated independently
(and according to Eqs.~(\ref{eq:38}, \ref{do})) at each time-step $t$.
The corresponding thresholds ${\cal Q}^+_{a}$ and ${\cal
  Q}^-_{a}$ appear to be lower than (respectively) the ${\cal
  Q}^+_{f}$ and ${\cal Q}^-_{f}$; cf. Eq.~(\ref{eq:1}). For
example, ${\cal Q}^-_{f}=0.265$ and ${\cal Q}^-_{a}=0.245$ for
parameters of Figs.~\ref{fig_new_1} and \ref{fig_new_2}.

As Figs.~\ref{fig_new_8}--\ref{fig_new_11} show, the activity pattern
in the annealed situation is similar to the noisy, non-weighted
situation described in Section \ref{section_noise} (recall however
that the latter does not have sharp thresholds; the activity there
decays together with the magnitude of noise). In particular, this
similarity concerns the factorization into clusters
[cf. Fig.~\ref{fig_new_8} with Fig.~\ref{fig4}] and short-term versus
long-term activity [cf. Fig.~\ref{fig_new_10} with Fig.~\ref{fig4_1}].

\section{Conclusion}
\label{sec:conclusion}

We have studied mechanisms for self-sustaining collective activity in
networks using an activation model inspired by neuronal dynamics
(\ref{n1}, \ref{n2}). Our specific set-up was motivated by the
empirical observation [see Section \ref{choice}] that the social
networks composed of functional links do not have long loops
(involving more then three nodes). As a concrete implementation of
this type of network, we focused on the Cayley tree with randomly
added triangles, and examine memory-induced mechanisms of sustaining
activity patterns on those networks.

We uncovered several scenarios where the network is capable of sustaining collective activity patterns for arbitrarily long times. For the non-weighted Cayley tree (with noiseless agents), we observe a fragmentation of the network into several clusters (see Section
\ref{sus}), so that the activity of the agents within each cluster is synchronized. The
clusterization and synchronization proceed along different
timescales: the former is (much) quicker than the latter.  It is thus
possible that at certain observation times, only clustering will be
observed. 

The collective activity sustaining is observed whenever the
inter-agent coupling $q$ is larger than a certain threshold ${\cal
  Q}^+$. Among other parameters, ${\cal Q}^+$ depends on the amount of
activation provided initially. The structure (and number) of clusters
depends not only on this amount, but also on which agents are
activated initially. For ${\cal Q}^-<q<{\cal Q}^+$ (where ${\cal Q}^-$
is a lower threshold), the dynamics strongly depend on initial
conditions. Thus, this model does show selectivity with respect to
initial activation, i.e., some agents play a role of effective activity
sustainers. Note that these features of activity sustaining thresholds
differ from those of cascade thresholds that depend mostly on the
network structure, even if the feedback effects on cascades are
accounted for \cite{arenas}.

The above dependencies on initial conditions are eliminated under the
influence of a (behavioral) agent noise; see Section
\ref{section_noise}. Under this noise, the network robustly fragments
into few (short-time synchronized) clusters, while the activity
sustaining does not have a threshold character. These conclusions do
not depend on the model of noise.

We also studied a more realistic situation where the network is
randomly weighted, i.e., there is a random distribution of
priorities. Our results indicate that the presence of weights does
facilitate the activity sustaining, and leads to a different scenario
of clusterization, where clusters are (physically) more isolated from
each other; see Section \ref{section_weighted}.

This study was confined to a model-dependent situation; in
particular, we worked on the level of the Cayley tree for the network
(not a fully realistic functional network), and we did not attempt any
direct data-fitting. But the results of this model do suggest several
directions for empiric (data-driven) research. To what extent can the
activity pattern in real social media be modeled via (partially
synchronized) clusters of agents? Do behavioral noise and
weighting (attention re-distribution) have to be accounted for explicitly?

We also note that in this study we assumed that all the connections are
facilitating; see Eq.~(\ref{eq:3}). It is known that the inhibitory
connections do play a crucial role in sustaining and shaping the
(biological) neuronal activity \cite{peto}. As future work, it will
be interesting to see the extent to which such inhibitory connections are
present in the interactions in social media, and how they affect the
collective activity patterns.

\acknowledgments This research was supported in part by DARPA grant
No. W911NF--12--1--0034.

%\clearpage

\clearpage

%%%% begin figures

%%%% 2

%%%% 2

\begin{figure*}[htbp]
    \includegraphics[width=0.6\columnwidth]{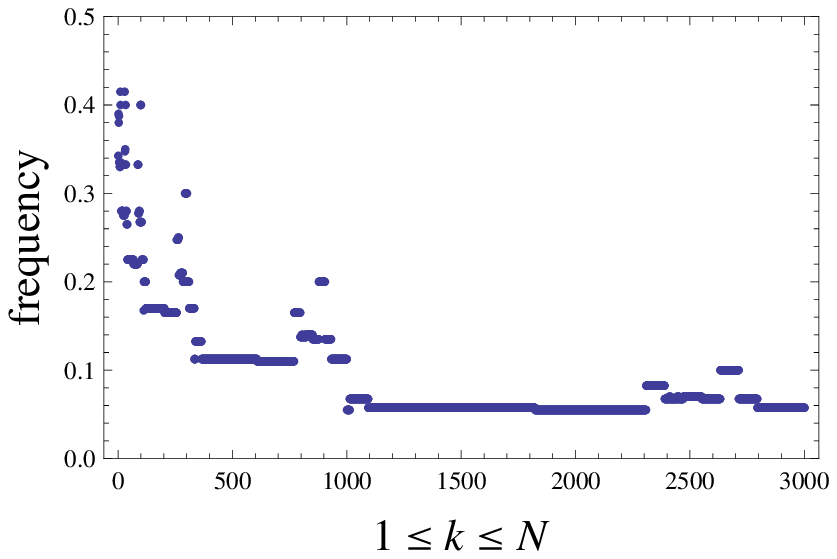} 
    \caption{The same as in Figs.~\ref{f1_a},
      but for another realization of the random initial condition
      generated with $b=1.8$; see Eq.~(\ref{eq:29}). }
\label{f2}
\end{figure*}

\comment{
\begin{figure*}[htbp]
\centering
\subfigure[]{
     \label{f2_a}
    \includegraphics[width=0.65\columnwidth]{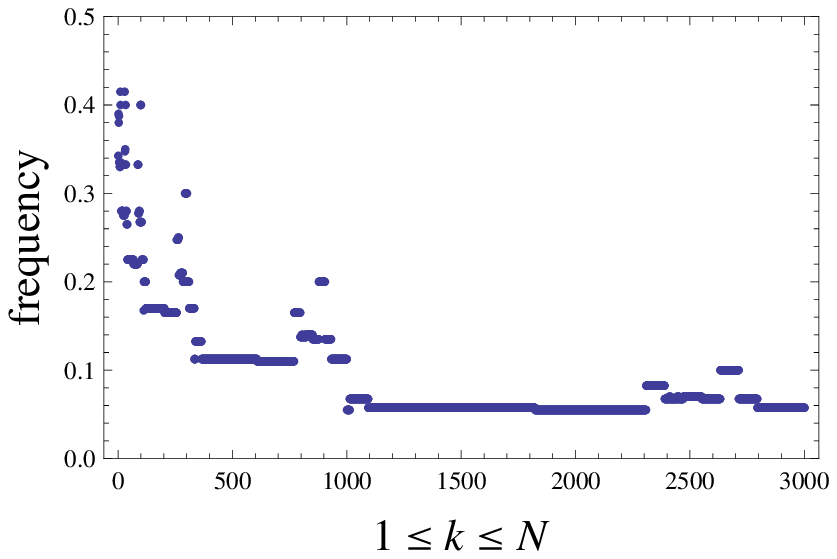} 
    } %\hspace*{1cm}
    \subfigure[]{
     \label{f2_b}
    \includegraphics[width=0.65\columnwidth]{bx_f1.eps} 
    }
    \subfigure[]{
     \label{f2_c}
   \includegraphics[width=0.65\columnwidth]{bx_f2.eps} 
    }
    \caption{The same as in Figs.~\ref{f1} (resp. (a), (b) and (c)),
      but for another realization of the random initial condition
      generated with $b=1.8$; see Eq.~(\ref{eq:29}). Here (b) and (c)
      refer respectively to the largest cluster in the periphery (with
      coordinates $1100\leq k\leq 2300$), and to the largest cluster
      in the core ($350\leq k\leq 780$).}
\label{f2}
\end{figure*}
}

\begin{figure*}[htbp]
\centering
\subfigure[]{
     \label{f3_a}
    \includegraphics[width=0.6\columnwidth]{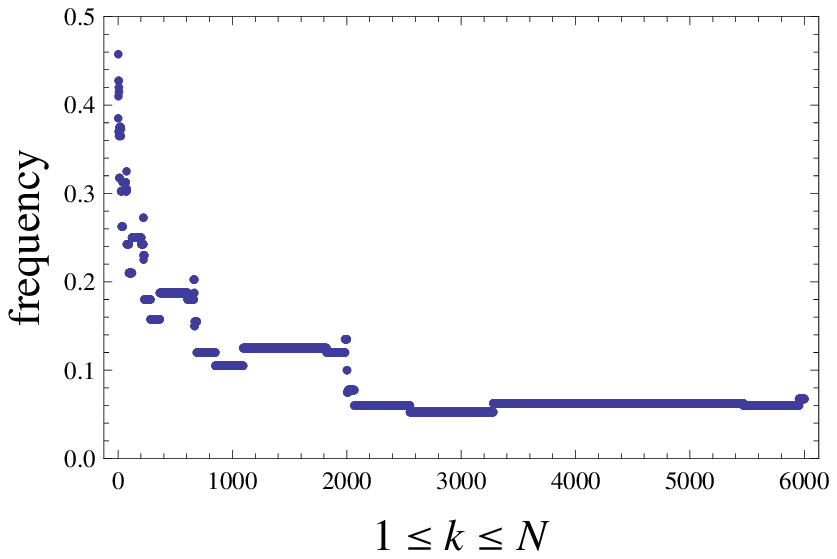} 
    } %\hspace*{1cm}
    \subfigure[]{
     \label{f3_b}
    \includegraphics[width=0.6\columnwidth]{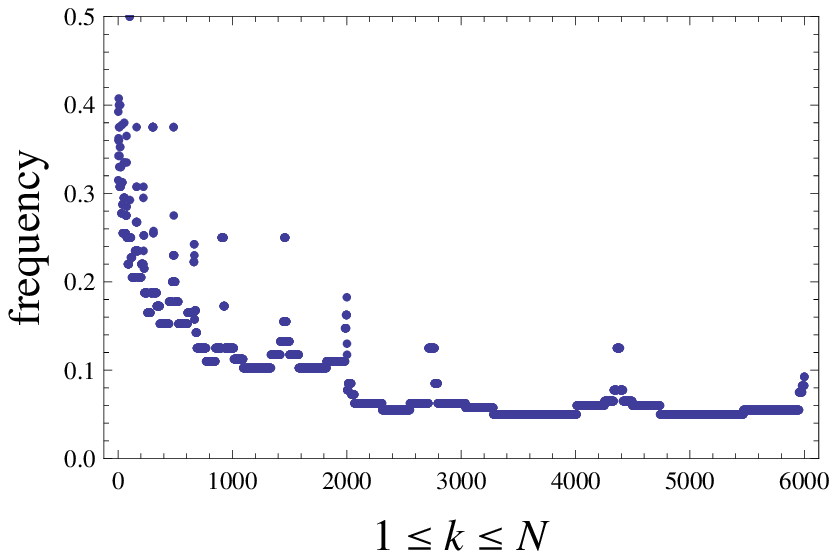} 
    }
    \caption{(Color Online) Comparison of firing frequencies with and
      without triangles. \\
      (a) The Cayley tree without triangles: frequency of fairing in
      the interval $[1200,1600]$ divided over the interval length
      $400$. The parameters are the
      same as in Figs.~\ref{f1}, but with $N=6\times 10^3+1$.\\
      (b) The same as in (a), but with randomly added 1050 links that
      define 1050 triangles.  }
     \label{f3}
\end{figure*}

%%% 5

\begin{figure*}[htbp]
\centering
\subfigure[]{
     \label{fig4}
    \includegraphics[width=0.6\columnwidth]{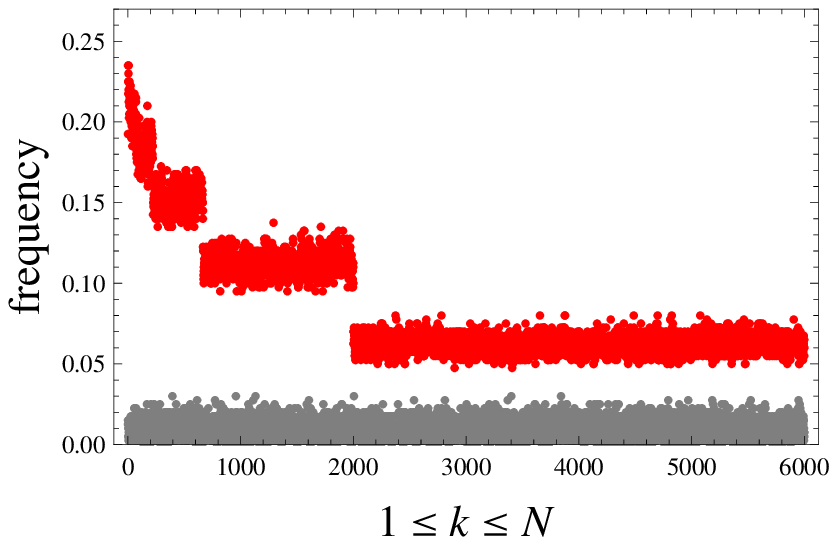} 
    } %\hspace*{1cm}
    \subfigure[]{
    \label{fig4_1}
    \includegraphics[width=0.6\columnwidth]{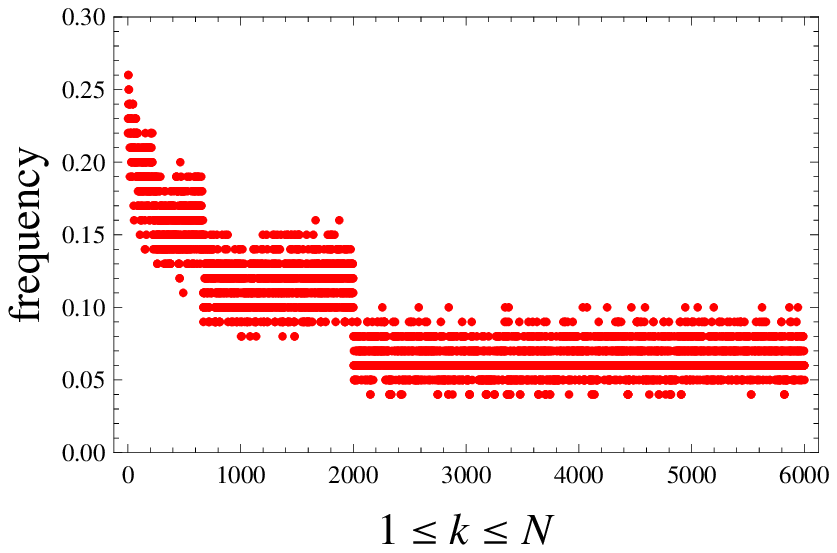}
    }    
    \caption{(Color Online) Dynamics under behavioral noise with
      magnitude $\eta=0.01$; cf. Eqs.~(\ref{noise}, \ref{n33}). \\
      (a) Red points (upper stripes): Frequency of each agent versus
      the agent number: the number of fairings in the interval
      $[400,800]$ divided over the interval length $400$ for $u=1$,
      $r=0.1$, $\kappa=0.101$, $K=3$, $N=6\times 10^3+1$ (total number
      of nodes), $q=0.3$ (subthreshold situation: $q<{\cal Q}^-$; see
      Table I). The initial state is passive: $w_i(0)=0$. The
      core-periphery border is at $k=2000$. The time-average of $m(t)$
      equals $0.08$. Gray points (the lowest stripe): the same as for
      blue points but for $q=0$ (isolated agents). The
      time-average of $m(t)$
      equals to the       noise magnitude $0.01$.  \\
      (b) The same as in (a), but the firings are counted in the
      interval $[400,500]$ and are divided over the interval length
      $100$.  }
\end{figure*}

\begin{figure*}[htbp]
\centering
    \subfigure[]{
    \label{fig4_2}
    \includegraphics[width=0.6\columnwidth]{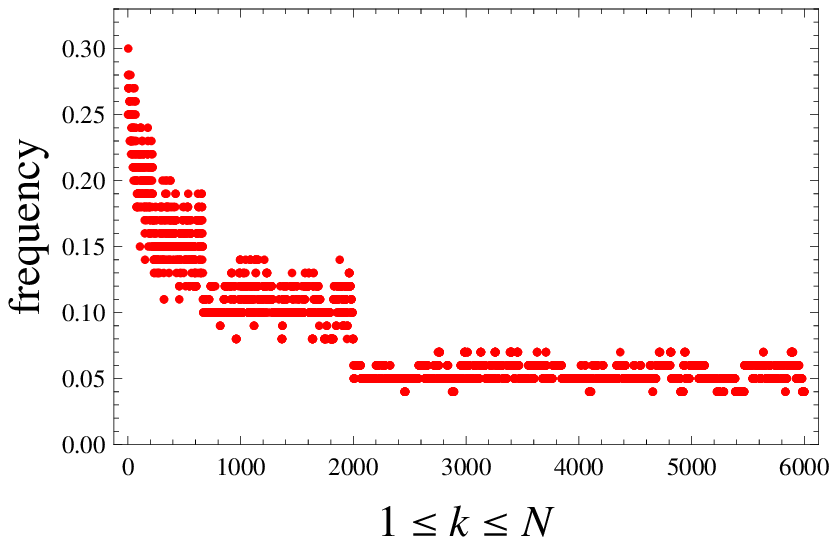}
    }  
\subfigure[]{
    \label{fig4_3}
    \includegraphics[width=0.6\columnwidth]{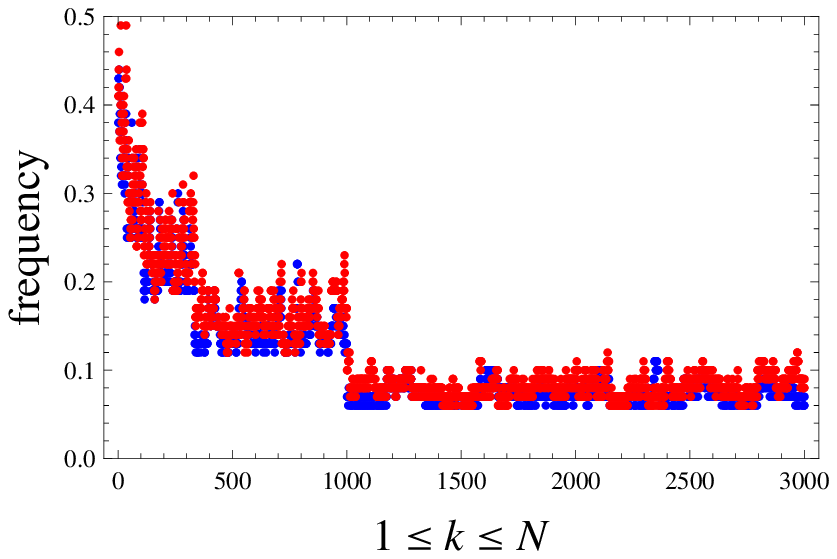}
    }
    \caption{(Color Online) (a) The same as in Fig.~\ref{fig4_1}, but
      the noise is implemented via
      Eq.~(\ref{noise_temperature}) with $\theta=0.01$. \\
      (b) Blue points: frequency in the time-interval $[0,100]$ for
      the same parameters as in Figs.~\ref{f1}. Red points: the same
      initial parameters and the same initial activity as for the blue
      points, but with noise $\eta=0.0001$; cf. Eqs.~(\ref{noise},
      \ref{n33}). The blue and red stripes largely overlap, but the
      red stripe is more narrow, hence the blue points border from
      below each stripe of red points. There is an approximate
      correspondence between the noisy and noiseless situations.  }
\end{figure*}

\begin{figure*}[htbp]
\centering
\subfigure[]{
    \label{fig7}
    \includegraphics[width=0.6\columnwidth]{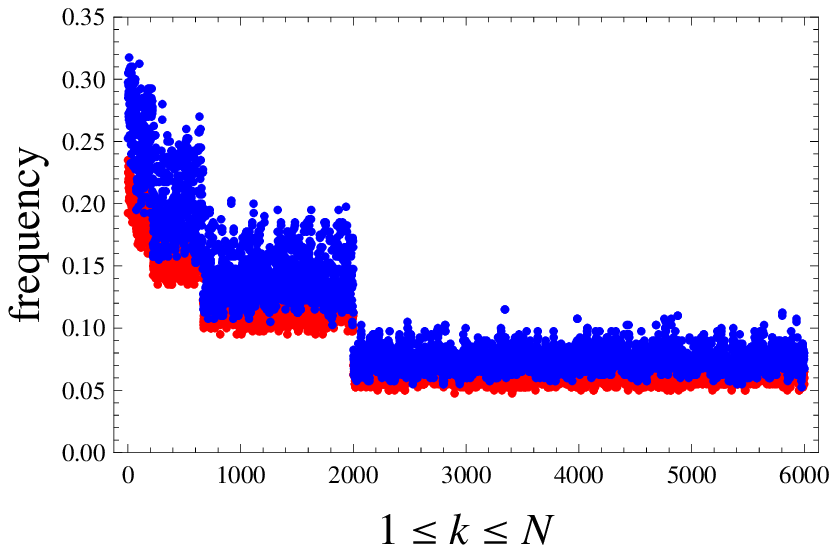}
    }
    \caption{(Color Online) Red points (lowest bands for each stripe)
      are the same as in Fig.~\ref{fig4}. Blue points (upper bands for
      each stripe): the same situation, but with randomly added 1050
      links that define 1050 triangles. For each stripe a narrow band
      of red points borders from below the wider band of blue
      points. The structure of clusters stays unchanged, but the
      activity increases after adding triangles.  }
\end{figure*}

%%%%% 5

\begin{figure*}[htbp]
\centering
\subfigure[]{
    \label{fig_new_1}
    \includegraphics[width=0.6\columnwidth]{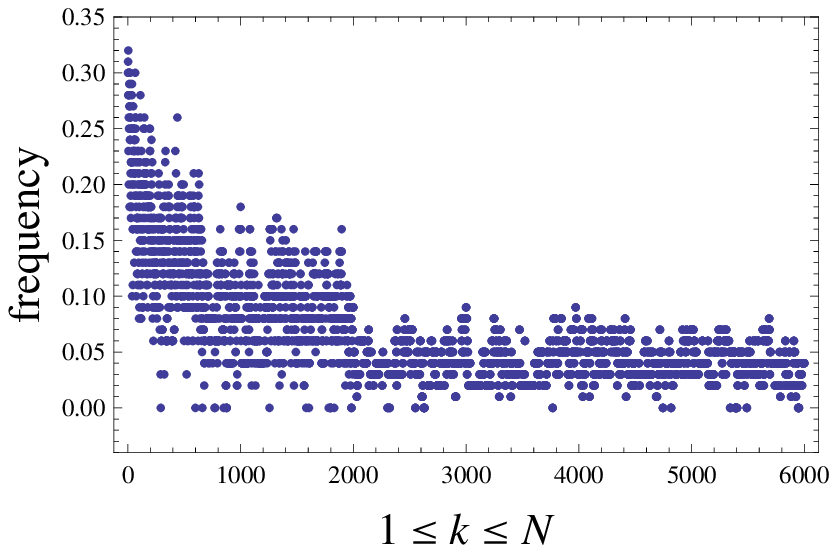} 
    } 
\subfigure[]{
    \label{fig_new_2}
    \includegraphics[width=0.6\columnwidth]{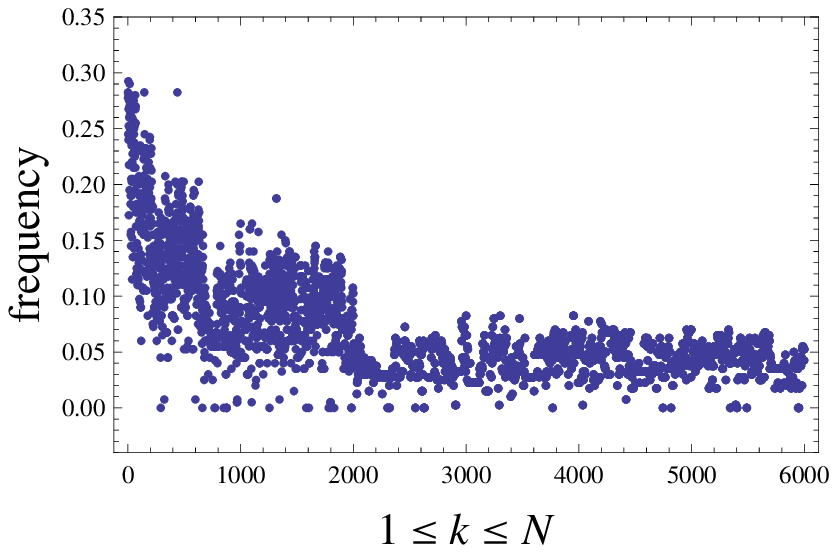} 
    } 
    \caption{(Color Online) Frozen random weights. For all figures
      $q=0.32$, $\kappa=0.101$, $u=1$, $r=0.1$, $K=3$, $N=6\times
      10^3+1$ (total number of agents). The initial condition was
      generated with $b=1.8$; see Eq.~(\ref{eq:29}). The random
      weights were
      generated with $b'=10$; see Eq.~(\ref{do}). \\
      (a) Short-time frequencies: the number of firing for each agent
      in the time-interval $[800,900]$ divided over the interval
      length $100$. \\
      (b) Long-time frequencies (in the same situation as (a)): the
      number of firing for each agent in the time-interval
      $[900,1300]$ divided over the interval length $400$.  }
\end{figure*}

\begin{figure*}[htbp]
\centering
    \includegraphics[width=0.6\columnwidth]{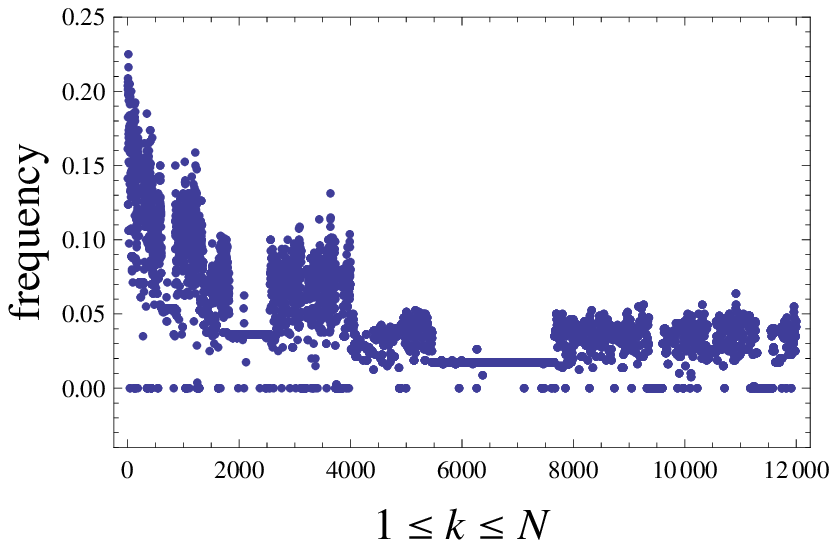} 
    \caption{(Color Online) Frozen random weights. For both figures
      $q=0.29$, $\kappa=0.101$, $u=1$, $r=0.1$, $K=3$, $N=12\times
      10^3+1$ (total number of agents). The initial condition was
      generated with $b=1.8$; see Eq.~(\ref{eq:29}). The random
      weights were generated with $b'=10$; see Eq.~(\ref{do}).
      Frequencies: the number of firing for each agent in the
      time-interval $[800,1600]$ divided over the interval length
      $800$.  }
    \label{fig_new_44}
\end{figure*}

\begin{figure*}[htbp]
\centering
\subfigure[]{
    \label{fig_new_8}
    \includegraphics[width=0.6\columnwidth]{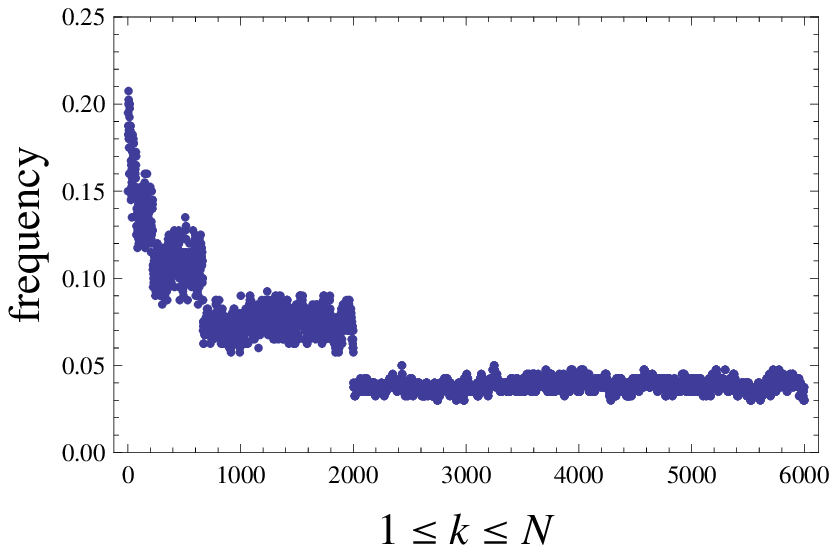} 
    } 
    \subfigure[]{
      \label{fig_new_9}
    \includegraphics[width=0.6\columnwidth]{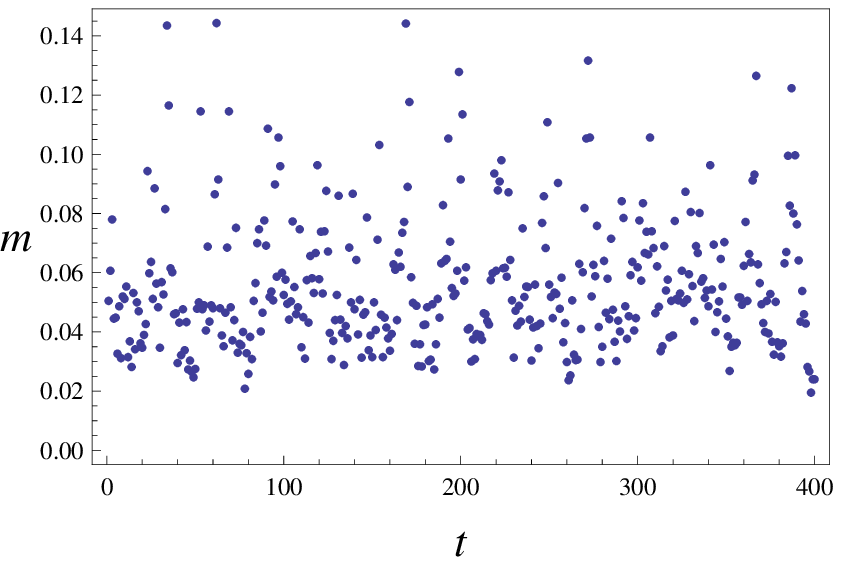} 
    }
    \caption{(Color Online) Annealed random weights. For all figures
      $u=1$, $\kappa=0.101$, $r=0.1$, $K=3$, $N=6\times 10^3+1$ (total
      number of agents). The initial condition was generated with
      $b=1.8$; see Eq.~(\ref{eq:29}). The random weights were
      generated with
      $b'=10$; see Eq.~(\ref{do}). \\
      (a) Frequency (long-time) of each agent versus the agent number
      for $q=0.32$. Frequencies are counted as the number of firing
      for each agent in the in the time-interval
      $[400,800]$ divided over the interval length $400$.    \\
      (b) The colective activity $m(t)=\frac{1}{N}\sum_{k=1}^N m_i(t)$
      versus discrete time $400+t$ for the same parameters as in
      (a). }
\end{figure*}

\begin{figure*}[htbp]
\centering
\subfigure[]{
  \label{fig_new_10}
    \includegraphics[width=0.6\columnwidth]{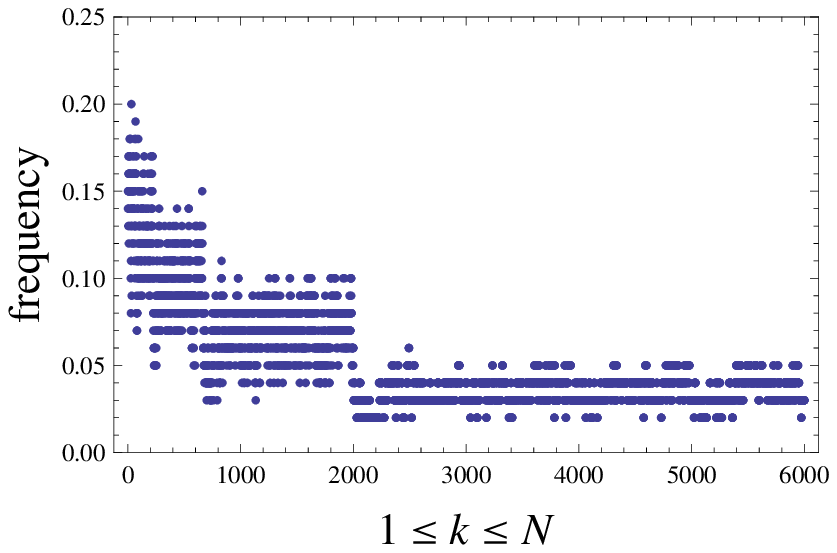} 
    }
\subfigure[]{
  \label{fig_new_11}
    \includegraphics[width=0.6\columnwidth]{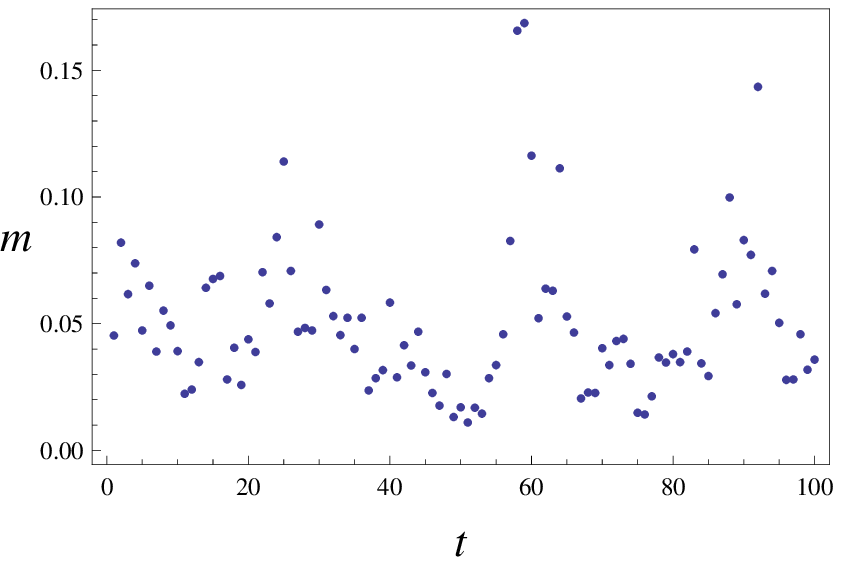} 
    }
    \caption{(Color Online)
The same parameters as in Figs.~\ref{fig_new_8} and
      \ref{fig_new_9}.\\
      (a) Frequency (short-time) of each agent versus the agent number
      $q=0.279$. Frequencies are counted as the number of firing for
      each agent in the in the time-interval
      $[300,400]$ divided over the interval length $100$. \\
      (b) The colective activity $m(t)=\frac{1}{N}\sum_{k=1}^N m_i(t)$
      versus discrete time $400+t$ for the same parameters as in
      (a). }
\end{figure*}

%%%% end figures


\begin{thebibliography}{99}


\bibitem{lazar}E. Katz and P.F. Lazarsfeld, {\it Personal Influence:
    The Part Played by People in the Flow of Mass Communication} (Free
  Press, Glencoe, Ill, 1955).


\bibitem{watts} D. J. Watts, PNAS {\bf 99}, 5766 (2002).

\bibitem{dodds_watts} 
P. S. Dodds and D. J. Watts, Phys. Rev. Lett. {\bf 92}, 218701 (2004).

%Memory in the threshold contagion model

\bibitem{galstyan} 
A. Galstyan and P. Cohen, Phys. Rev. E {\bf 75}, 036109 (2007).

%Influence of communites on cascading process

\bibitem{gleeson} J. Gleeson, Phys. Rev. E {\bf 77}, 046117 (2008).

% A. Hackett, S. Melnik, and J. Gleeson, Physical Review E 83, 056107
% (2011).

\bibitem{porter2014} M. A. Porter, J. P. Gleeson, {\it Dynamical
    Systems on Networks: A Tutorial}, arXiv:1403.7663. 

\bibitem{barrat} A. Barrat, M. Barth\'{e}lemy and A. Vespignani, {\it
    Dynamical Processes on Complex Networks} (Cambridge University
  Press, Cambridge, 2008).


\bibitem{weak_ties}
M. Granovetter, Am. J. Sociol. {\bf 83}, 1420 (1978).

% The strength of weak ties

\bibitem{granovetter} M. Granovetter and R. Soong,
  Sociol. Methodol. {\bf 18}, 69 (1988).

  % {\it Threshold models of diversity: Chinese restaurants, residential
  %   segregation, and the spiral of silence}

\bibitem{morris}
S. Morris, Rev. Econ. Stud. {\bf 67}, 57 (2000).

\bibitem{durlauf}
S. N. Durlauf, Sociological Methodology, {\bf 31}, 47 (2001).

% A framework for the study of individual behavior and social
% interactions

\bibitem{stauffer}
G. Weisbuch and D. Stauffer, Physica A
323, 651 (2003).

% Adjustment and social choice,

\bibitem{econo} S. Sinha, A. Chatterjee, A. Chakrabarti and
  B. K. Chakrabarti, {\it Econophysics: An Introduction} (Wiley-VCH,
  Weinheim, 2010).

\bibitem{cortes}
G.C. Galster, R.G. Quercia and A. Cortes, Housing Policy
Debate, {\bf 11}, 701 (2000). 

% Identifying neighborhood thresholds: An empirical exploration.



\bibitem{arbib}  {\it The Handbook of Brain Theory and Neural
    Networks}, ed. by M. A. Arbib (MIT Press, Cambridge, 2003).  

\bibitem{peto}
P. Peretto, {\it An introduction to the modelling of neuronal networks}
(Cambridge University Press, Cambridge, 1994).

\bibitem{stro}R.E. Mirollo and S.H. Strogatz, SIAM Journal on Applied
  Mathematics, {\bf 50}, 1645 (1990).

\bibitem{integrate_fire} W. Gerstner, {\it Integrate-and-Fire Neurons
    and Networks}, in {\it The Handbook of Brain Theory and Neural
    Networks}, ed. by M. A. Arbib (MIT Press, Cambridge, 2003).


\bibitem{klopf}A.H. Klopf, {\it The hedonistic neuron: a theory for
    learning, memory and intelligence} (Hemisphere, New York, 1982). 

\bibitem{shvyrkov}
  V.B. Shvyrkov, Advances in Psychology {\bf 25}, 47 (1985).

% Toward a psychophysiological theory of behavior

\bibitem{nowak} A. Nowak, R. R. Vallacher and E. Burnstein, {\it
    Computational social psychology: A neural network approach to
    interpersonal dynamics} in {\it Computer Modelling of Social
    Processes}, ed. by W.B.G. Liebrand, A. Nowak and R. Hegselmann
  (London, Sage, 1998).
  
\bibitem{vidal} J.M. Vidal and E.H. Durfee, {\it Multiagent
    systems}, in {\it The Handbook of Brain Theory and Neural
    Networks}, ed. by M. A. Arbib (MIT Press, Cambridge, 2003).

\bibitem{larson} L.J. Larson-Prior, {\it Parallels in Neural and Human
    Communication Networks} in {\it Handbook of Human Computation},
  ed. by P. Michelucci (Springer Science + Business Media, New York
  2013).


\bibitem{roxin} 
A. Roxin, H. Riecke and S. A. Solla, Phys. Rev. Lett. {\bf 92}, 198101
(2004).

%% Self-sustained activity in a small-world network of excitable neurons

\bibitem{kaski} S. Sinha, J. Saramaki and K. Kaski, Phys. Rev. E
{\bf 76}, 015101 (2007).

% Emergence of self-sustained patterns in small-world excitable media

\bibitem{qian} Y. Qian, X. Liao, X. Huang, Y. Mi, L. Zhang and G. Hu,
  Phys. Rev. E {\bf 82}, 026107 (2010).

% Diverse self-sustained oscillatory patterns and their mechanisms in
% excitable small-world networks.


\bibitem{arenas} P. Piedrahita, J. Borge-Holthoefer, Y. Moreno and
  A. Arenas, EPL, {\bf 104}, 48004 (2013).

% Modeling self-sustained activity cascades in socio-technical networks,


\bibitem{bassett} D.S. Bassett and E. Bullmore, The Neuroscientist
  {\bf 12}, 512 (2006).
%Small-world brain networks. 

\bibitem{davidsen}J. Davidsen, H. Ebel and S. Bornholdt, Phys. Rev.
  Lett. {\bf 88}, 128701 (2002).

% Emergence of a small world
%   from local interactions: mod- eling acquaintance networks.


\bibitem{E-Y}
L. B. Emelyanov-Yaroslavsky and V. I. Potapov, 
Biol. Cybern. {\bf 67}, 67 (1992); {\it ibid.} {\bf } 73 (1992).


\bibitem{anni}
  G.C. Garcia, A. Lesne, M.T. Hutt and C.C. Hilgetag,
 Frontiers in computational neuroscience, {\bf 6}, 50 (2012).

  % Building blocks of self-sustained activity in a simple deterministic
  % model of excitable neural networks


\bibitem{mcgraw}
P. McGraw and M. Menzinger, Phys. Rev. E {\bf 83}, 037102 (2011).

\bibitem{chin}
 Z. Li-Sheng, G. Wei-Feng, H. Gang, and M. Yuan-Yuan, Chin. Phys. B
 {\bf 23}, 108902
 (2014).

 % Network dynamics and its relationships to topology and coupling
 % structure in excitable complex networks


\bibitem{abbott} C.  Vanvreeswijk and L.F. Abbott, SIAM Journal on Applied
 Mathematics, {\bf 53}, 253 (1993).

 % Self-sustained firing in populations of integrate-and-fire neurons.

\bibitem{schuster} M. Usher, H.G. Schuster and E. Niebur,
  Neural Computation, {\bf 5}, 570 (1993).

  % Dynamics of populations of integrate-and-fire neurons, partial
  % synchronization and memory


\bibitem{fries}P. Fries, Trends Cogn. Sci. (Regul. Ed.) {\bf 9}, 474
  (2005).

% A mechanism for cognitive dynamics: neuronal communication through
% neuronal coherence.

\bibitem{pipa} P. J. Uhlhaas, G. Pipa, B. Lima, L. Melloni,
  S. Neuenschwander, D. Nikolic and W. Singer,
  Front. Integr. Neurosci. {\bf 3}, 17 (2009).

  % Neural synchrony in cortical networks: history, concept and current
  % status



\bibitem{mikhailov} S.C. Manrubia, A.S. Mikhailov and D.H. Zanette,
  {\it Emergence of dynamical order} (World Scientific, Singapore, 2004).


\bibitem{others}
G. Bub, A. Shrier, and L. Glass. Phys. Rev. Lett. {\bf 94}, 028105
(2005).

A.J. Steele, M. Tinsley and K. Showalter, Chaos {\bf 16}, 015110 (2006).

%% Spatiotemporal dynamics of networks of excitable nodes

% M.R. Tinsley, A.F. Taylor, Z. Huang, F.Wang, K. Showalter, Physica
% D {\bf 239}, 785 (2010).


\bibitem{kaneko} K. Kaneko, Physica D {\bf 41}, 137 (1990).

\bibitem{crisanti}A. Crisanti, M. Falcioni and A. Vulpiani,
  Phys. Rev. Lett. {\bf 76}, 612 (1996).

\bibitem{manrubia_non_ergo} S. C. Manrubia and A. S. Mikhailov, EPL,
  {\bf 53}, 451 (2001).


\bibitem{gade}P. M. Gade, H. A. Cerdeira, and R. Ramaswamy,
  Phys. Rev. E {\bf 52}, 2478 (1995).

  \bibitem{mimi}A.S. Mikhailov and V. Calenbuhr, {\it
From Cells to Societies} (Springer, Berlin, 2002).

\bibitem{huberman}
B. A. Huberman, D. M. Romero, and F. Wu, {\it Social networks that matter:
Twitter under the microscope}, arXiv:0812.1045v1. 

% What is Twitter, a Social Network or a News Media?
% Haewoon Kwak, Changhyun Lee, Hosung Park, and Sue Moon


\bibitem{www2012} G. Ver Steeg and A. Galstyan, {\it Information
    Transfer in Social Media}, in Proceedings of World Wide Web
  Conference (WWW), Lyon, France, 2012.

\bibitem{wsdm2013} G. Ver Steeg and A. Galstyan, {\it
    Information-Theoretic Measures of Influence Based on Content
    Dynamics}, in Proceedings of WSDM'13, Rome, Italy, 2013.


\bibitem{malsburg}
C. von der Malsburg, Kybernetik, {\bf 14}, 85 (1973).



\bibitem{hu_at}B. Huberman, F. Wu, {\it The Economics Of Attention:
    Maximizing User Value In Information-Rich Environments},
  Proceedings of the 1st international workshop on Data mining and
  audience intelligence for advertising (ADKDD-07), 16-20 (2007).

\bibitem{v_at} B. Goncalves, N. Perra, A. Vespignani,   
PLOS ONE {\bf 6}, e22656 (2011).

% Modeling Users Activity on Twitter Networks: Validation of Dunbar's
% Number.


\bibitem{latane}
B. Latane, American Psychologist, {\bf 36}, 343
  (1981). % The psychology of social impact.

\bibitem{freeman}L.C. Freeman, Social Networks, {\bf 1}, 215 (1979).

\bibitem{everett} S. P. Borgatti and M. G. Everett, Social Networks,
  {\bf 21} 375 (1999).

%% Models of core-periphery structure, pp 375 -- 395

\bibitem{bocca}
S. Boccaletti, V. Latora, Y. Moreno, M. Chavez, D.-U. Hwang, Phys. Rep. {\bf 424}, 175 (2006).

%% Complex networks: Structure and dynamics

\bibitem{puck}
M. P. Rombach, M. A. Porter, J. H. Fowler, P. J. Mucha
SIAM Journal of Applied Mathematics {\bf 74}, 167 (2014).

%% Core-Periphery Structure in Networks


%% synch

\bibitem{motter}
 C. Zhou, A. E. Motter and J. Kurths, Phys. Rev. Lett. {\bf 96},
 034101 (2006). 

  % Universality in the Synchronization of Weighted Random Networks

  
\bibitem{touboul} G. Hermann and J. Touboul, Phys. Rev. Lett. {\bf 109},
  018702 (2012).

  
\bibitem{peron} T.D.K.M. Peron, P. Ji, F. A. Rodrigues, and J. Kurths,
  {\it Impact of order three cycles in complex network spectra},
  arXiv:1310.3389.

\bibitem{binder} 
K. Binder and A. P. Young, Rev. Mod. Phys. {\bf 58}, 801 (1986).

% Spin glasses: Experimental facts, theoretical concepts, and open
% questions



\end{thebibliography}
\end{document}